\documentclass[aps,prd,twocolumn,superscriptaddress,nobibnotes]{revtex4-2}
\usepackage{amsmath}
\usepackage{amsfonts}
\usepackage{amssymb}
\usepackage{multirow}
\usepackage{slashed}
\usepackage{tikz}
\usetikzlibrary{shapes,arrows,positioning,decorations.pathmorphing,patterns,calc}    
\usepackage{booktabs}
\usepackage{graphicx} 

\usepackage{dcolumn}

\usepackage{mathtools}
\usepackage[colorlinks,
linkcolor = green,
urlcolor  = blue,
citecolor = blue,
anchorcolor = blue]{hyperref}
\begin{document}
\title{Spectroscopic Properties of the Molecular \( T_{cc}^{+} \) Meson in a Thermal Medium}
	
	\author{ S. Damen}
	\affiliation{Department of Physics, Kocaeli University, 41001 Izmit, T\"urkiye}
	
	\author{J.Y.~S\"ung\"u}
	\affiliation{Department of Physics, Kocaeli University, 41001 Izmit, T\"urkiye}
	\author{E. Veli Veliev}
	\affiliation{Department of Physics, Kocaeli University, 41001 Izmit, T\"urkiye}
\begin{abstract}
In this work, we investigate the exotic doubly charmed molecular state $T_{cc}^{+}(3875)$ with quantum numbers $J^{P} = 1^{+}$ using the Thermal QCD Sum Rules framework. Employing a molecular interpolating current, we evaluate the two-point correlation function by incorporating non-perturbative condensate contributions up to dimension six. From the resulting thermal sum rules, we determine the temperature dependence of the mass, decay constant, and width of the $T_{cc}^{+}$ state. Our numerical analysis reveals that all quantities remain remarkably stable under temperature variations up to $T \simeq 120~\text{MeV}$, after which they change significantly. At the deconfinement temperature, the mass decreases to approximately $28\%$ of its vacuum value, and the decay constant drops to about $25\%$. We analyze the thermal evolution of the decay width of $T_{cc}^{+}$, finding $\Gamma_{T_{cc}^{+}}(0) = 434.95 \pm 7.66~\text{keV}$ at zero temperature. The width of $T_{cc}^{+}$ remains unchanged until $ T\simeq 120~\text{MeV}$, after which it begins to grow rapidly. The investigation of thermal effects on $T_{cc}^{+}$ provides new insights into QCD phase transitions, chiral symmetry restoration, and the properties of strongly interacting, hot, and dense matter. These findings are expected to serve as useful input for future experimental searches and phenomenological studies of exotic mesons.
\end{abstract}
\maketitle
\section{Introduction}\label{sec:Int}
Over the past two decades, numerous exotic hadrons have been experimentally discovered, rapidly expanding the so-called particle zoo. Among these, four-quark states have attracted considerable attention since the discovery of the $ X(3872)$. Within the framework of Quantum Chromodynamics (QCD), several scenarios are possible, including compact tetraquarks in a diquark-antidiquark structure, loosely bound molecular states in meson-meson configurations, hadrocharmonium, adjoint charmonium, and others. Despite intense theoretical and experimental efforts, there is still no compelling consensus on the internal nature of these exotic states, including the broader family of $\text{XYZ}$ mesons \cite{Belle:2003nnu, Chen:2016qju, Esposito:2016noz, Guo:2017jvc}.

Doubly charmed four-quark systems have also been investigated within the framework of the hadronic molecular picture, in which they are modelled as bound states of conventional mesons. It is worth emphasizing that the study of charmonium-like molecular structures is not a recent development; such systems were already considered in the literature several years ago \cite{Deng:2021gnb,Agaev:2022ast,
Albaladejo:2021vln,Wang:2022jop,Feijoo:2021ppq,Agaev:2022vhq,Braaten:2020nwp,Cheng:2020wxa,Ferah:2025jle,Azizi:2021aib}. Many theoretical studies of the doubly heavy tetraquark states are based on the compact diquark-antidiquark picture through various theoretical methods, such as various quark models \cite{Meng:2021yjr, Qin:2020zlg, Richard:2018yrm, Yan:2018gik, Lin:2024gcm},  coupled-channel framework~\cite{Kim:2025ado} and QCD sum rules (QCDSR) \cite{Navarra:2007yw,Dias:2011mi,Du:2012wp,Veliev:2011kq,Veliev:2008gx,Aydin:2025lbl,Veliev:2008zi,Sundu:2016oda,Agaev:2016urs,Agaev:2019qqn,Agaev:2022ast,Agaev:2023ruu}.

It is now well established that some hadrons may, in fact, be molecular states emerging from the interaction of more elementary hadrons, particularly in the charm sector. Of special significance for the field was the discovery, in the summer of 2021, of the charged exotic state $ T_{cc}^{+}$ by the LHCb Collaboration \cite{LHCb:2021vvq,LHCb:2021auc}. This very narrow state, observed in the $ D^{0} D^{0}\pi^{+}$ mass distribution, has a mass of  $m_{\text{thr}} = 3874.74 \pm 0.1~\text{MeV}$, corresponding  to the~$ D^{*+} D^{0}$ threshold. The measured width is $\Gamma = 410\pm 165\pm 43^{+18}_{-38}~\text{keV}$~\cite{LHCb:2021auc,ParticleDataGroup:2024cfk,Muheim2021}. Moreover, experimental results indicate that the spin–parity quantum numbers of the isoscalar $ T_{cc}^{+}$ favor an assignment of $ J^{P} = 1^{+}$. Significant disagreements remain regarding the internal structure of the doubly charmed $ T_{cc}^{+}$. Two main interpretations are typically considered for this phenomenon. Since its mass lies very close to the $ D^{*+} D^{0}$ threshold, its interpretation as a doubly charmed hadronic molecular state has been extensively discussed. Nevertheless, the possibility that it is a genuine doubly charmed tetraquark cannot be excluded. In this context, the properties of $ T_{cc}^{+}$ have also been investigated within the diquark–antidiquark and compact tetraquark scenarios \cite{Weng:2021hje,Kim:2022mpa}. 
\begin{figure}[h!]
    \centering
    \begin{tikzpicture}[scale=0.8]
        
        \definecolor{mesonblue}{RGB}{79, 129, 189}
        \definecolor{mesonpink}{RGB}{192, 80, 77}
        \definecolor{quarkc}{RGB}{218, 150, 148}
        \definecolor{quarku}{RGB}{155, 187, 89}
        \definecolor{quarkd}{RGB}{128, 158, 194}
        
        \draw[fill=mesonpink!20, thick, rounded corners=5pt] (-4, 1.5) rectangle (-2, 3.5);
        \node[font=\Large\bfseries] at (-3, 3.8) {$D^0$};
        \node[font=\footnotesize] at (-3, 1.2) {$J^P = 0^-$};
        
        \shadedraw[ball color=quarkc, draw=black] (-3.5, 2.5) circle (0.3cm);
        \node[font=\bfseries, white] at (-3.5, 2.5) {$c$};
        
        \shadedraw[ball color=quarku, draw=black] (-2.5, 2.5) circle (0.3cm);
        \node[font=\bfseries, white] at (-2.5, 2.5) {$u$};
        \draw[line width=1pt, white] (-2.7, 2.65) -- (-2.3, 2.65);
        
        \draw[fill=mesonblue!20, thick, rounded corners=5pt] (2, 1.5) rectangle (4, 3.5);
        \node[font=\Large\bfseries] at (3, 3.8) {$D^{*+}$};
        \node[font=\footnotesize] at (3, 1.2) {$J^P = 1^-$};
        
        \shadedraw[ball color=quarkc, draw=black] (2.5, 2.5) circle (0.3cm);
        \node[font=\bfseries, white] at (2.5, 2.5) {$c$};
        
        \shadedraw[ball color=quarkd, draw=black] (3.5, 2.5) circle (0.3cm);
        \node[font=\bfseries, white] at (3.5, 2.5) {$d$};
        \draw[line width=1pt, white] (3.3, 2.65) -- (3.7, 2.65);
        
        \draw[<->, line width=2pt, mesonblue] (-1.8, 2.5) -- (1.8, 2.5);
        \node[fill=white, rounded corners=3pt, inner sep=3pt, font=\small\bfseries] at (0, 2.5) {Molecular Binding};
        
        \draw[decorate, decoration={snake, amplitude=0.5mm, segment length=2mm}, mesonpink, thick] 
            (-3.2, 2.5) -- (-1.5, 2.5);
        \draw[decorate, decoration={snake, amplitude=0.5mm, segment length=2mm}, mesonblue, thick] 
            (3.2, 2.5) -- (1.5, 2.5);
        
    \end{tikzpicture}

    \caption{%
    Hadronic molecule structure of $T^{+}_{cc}$: bound state of $ D^{*+}$ and $ D^{0}$ mesons.%
    }
    \label{fig:TccMolecule}
\end{figure}
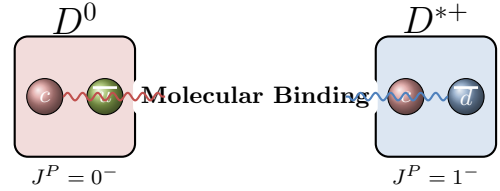\\

Heavy-quark spin symmetry predicts an isoscalar $D^{*}D^{*}$ molecular partner with $J^{P}=1^{+}$ \cite{Du:2021zzh}. Coupled-channel analysis of the $D^{*}D$ system has been motivated by recent LHCb results that reported a new state observed in the $D^{0}D^{0}\pi^{+}$ spectrum. The $T_{cc}^{+}$ appears as a bound state in the $D^{*}D$ amplitude with a dominant molecular composition \cite{Albaladejo:2021vln}.

The $T_{cc}^{+}$ is thus identified as an isoscalar state, and its properties are examined within a hadronic-molecular framework. The hadronic-molecular framework couplings of equal magnitude but opposite signs to the $D^{*+}D^{0}$ and $D^{+}D^{*0}$ channels. However, the wave function is dominated by the $D^{*+}D^{0}$ component due to the mass proximity to this channel's threshold \cite{Du:2021zzh}.

 In particular, the axial-vector state $ cc\bar{u}\bar{d}$, interpreted as the hadronic molecule $ M_{cc}^{+} \equiv  D^{0} D^{*+}$ composed of the ordinary mesons $ D^{0}$ and $ D^{*+}$, was analyzed in Refs.~\cite{Dias:2011mi,Agaev:2022ast}. The mass of $ M_{cc}^{+}$ was estimated in Ref.~\cite{Dias:2011mi} using the QCDSR method. The obtained prediction, $ m_{ M_{cc}^{+}} = (3872.2 \pm 39.5)~\text{MeV}$, indicates that this molecule is kinematically forbidden to decay into $ D^{0}$ and $ D^{*+}$ mesons. However, its mass is sufficiently large to allow the strong decay channel $ M_{cc}^{+} \to  D^{0} D^{0}\pi^{+}$. The predicted mass \cite{Agaev:2022ast}, $ m = (4060 \pm 130)~\text{MeV}$, and width, $\Gamma = (3.8 \pm 1.7)~\text{MeV}$, are larger than the corresponding values reported by the LHCb Collaboration.

Additionally, the behavior of the exotic $ T_{cc}^{+}$ state in a nuclear medium is investigated, with emphasis placed on the finite-density regime relevant to heavy-ion collisions at CBM (FAIR) in Ref.~\cite{Montesinos:2023qbx}. Within a theoretical framework based on the principles of heavy-quark spin symmetry (HQSS), the $T_{cc}^{+}$ is characterized as a bound state dynamically produced from the leading-order interaction between $DD^{*}$. The existence of its partner, the $ T_{cc}^{*}(4016)^{+}$, associated with the $D^{*}D^{*}$ channel, is further predicted by the HQSS. Generalized interactions beyond the purely molecular scenario are also explored, allowing for mixed configurations. Medium-induced modifications of the mass and width of $T_{cc}^{+}$ and $ T_{cc}^{*}$ are demonstrated \cite{Montesinos:2025mfx}, providing theoretical predictions that can be tested in forthcoming heavy-ion collision experiments at CBM.

Furthermore, the study of strongly interacting matter under extreme conditions remains a central challenge in the Thermal Sum Rules (TQCDSR). At sufficiently high temperatures and/or densities, quarks and gluons are expected to deconfine, forming a novel state of matter known as Quark-Gluon Plasma (QGP) \cite{Zhou:2025bwu,Brandt:2025sut,ALICE:2008ngc,Matsui:1986dk}. In this hot and dense environment, the chiral phase transition is anticipated to occur at the characteristic critical temperature. However, the precise quantitative description of deconfinement and chiral symmetry restoration is still incomplete, making this a vibrant area of ongoing research. The phase structure of the QGP encodes essential information about strong interactions in thermal media and offers insights into confinement mechanisms, hadronization, QCD vacuum dynamics, the physics of compact stars, and even the evolution of matter in the early universe~\cite{Chernodub:2024wis,Sungu:2020zvk}.

Experimental investigations, particularly through ultrarelativistic heavy-ion collisions at RHIC, LHC, and future facilities such as FAIR and NICA, recreate conditions similar to those of the early universe. These experiments provide crucial data for modeling hot and dense QCD matter, which is essential for mapping the QCD phase diagram—one of the primary goals of modern high-energy physics.
 A precise determination of the critical temperature \(  T_c \) marking the transition from hadronic matter to QGP would represent a milestone in understanding QCD under extreme conditions \cite{Chen:2024aom}.

In particular, thermal media induce characteristic modifications to meson properties, such as mass reductions driven by condensate depletion, enhanced decay widths arising from stronger interactions with the medium, and suppressed coupling constants reflecting weaker meson-quark binding. These variations encode vital information about non-perturbative dynamics and provide essential input for interpreting observables in high-energy collisions. Theoretical models incorporating thermal effects predict observable changes in meson spectral functions, such as residue suppression and width broadening, which serve as clear indicators of in-medium dissociation~\cite{Barata:2025jhd,Montana:2023sft}.

The study of exotic hadrons under thermal stress provides further opportunities to probe QCD matter. Compact tetraquarks with stronger binding are expected to be more resistant to dissociation than loosely bound molecular states, which tend to dissolve at lower temperatures. Heavy-ion collisions thus offer a unique testing ground for exotic candidates such as \(  T_{cc}^{+} \). Detecting or constraining their survival probabilities in thermal environments would not only clarify their internal structure but also strengthen our understanding of QCD matter at finite temperature and density~\cite{Wu:2020zbx}. Motivated by the above considerations, in this work we employ the TQCDSR 
framework to systematically investigate the thermal spectroscopic properties 
of the $T_{cc}^{+}$ state and its bottom-sector counterpart $T_{bb}^{+}$.

This manuscript is structured as follows: Section~\ref{sec:MassQCDSUMRULES} outlines the theoretical foundation of the TQCDSR methodology and its use for the \(  T_{cc}^{+} \) state and its b-partner, providing analytical expressions that determine the mass, decay constant and width up to dimension six operators. In Sec.~\ref{sec:Numerical}, we carry out the numerical analysis and report the corresponding results. 
Finally, Sec.~\ref{sec:conclusions} summarizes our main findings and conclusions.
\section{TQCDSR Formalism: Two-Point Correlator} \label{sec:MassQCDSUMRULES}
The QCDSR method is a well-established nonperturbative technique that has been extensively applied to the investigation of hadronic mass spectra and decay properties \cite{Shifman:1978bx,Reinders:1984sr}. To examine the temperature dependence of the mass, decay constant and width of the axial-vector \( T_{cc}^{+} \) state, we extend this framework to finite temperature using the TQCDSR formalism. In this section, we derive the QCDSR for the mass, decay constant, and decay width of the \( T_{cc}^{+} \) resonance. Our analysis closely follows the methods described in for both zero~\cite{Azizi:2008tt, Gungor:2023ksu,Azizi:2019abc,Sungu:2018eej,Navarra:2000ji}, and finite-temperature scenarios~\cite{Ayala:2016vnt, Mallik:1997kj, Mallik:1997pq,Dominguez:2012um,Dominguez:2007}. We begin by defining the thermal two-point correlation function \cite{Bochkarev:1985ex}:
\begin{equation}\label{eq:CorrF1}
\Pi_{\mu\nu}(q,T) = i \int d^{4}x\, e^{iq \cdot x} \langle w |
\mathcal{T} \{ J_{\mu}(x) J_{\nu}^{\dagger}(0) \} | w\rangle,
\end{equation}
here $\mathcal{T}$ denotes the time-ordering operator for currents, $|w\rangle$ represents the thermal medium at a temperature $T$, and $ J_\mu(x)$ is the interpolating current for the state $T_{cc}^{+}$. The thermal expectation value of an operator $A$ in equilibrium is defined as: 
\begin{equation}
\langle A \rangle = \frac{\mathrm{Tr}(e^{-\beta H}A)}{\mathrm{Tr}(e^{-\beta H})},
\end{equation}
where $H$ is the QCD Hamiltonian, $\beta=1/T$ is the inverse temperature. 

To construct the physical side of the sum rules, the correlation function is saturated by a complete set of intermediate hadronic states sharing the same quantum numbers $ J^{P} = 1^{+}$ as the $ T_{cc}^{+}$. Performing the Fourier transform leads to:
\begin{equation}\label{eq:phys1}
\Pi^{Phys}_{\mu\nu}(q,T) = \frac{\langle w | J_{\mu} |T_{cc}^{+}(q) \rangle \langle T_{cc}^{+}(q) | J_{\nu}^{\dagger} | w \rangle}{m_{T_{cc}^{+}}^{2}(T) - q^{2}} + \cdots,
\end{equation}
where $m_{T_{cc}^{+}}(T)$ is the temperature-dependent mass of the $T_{cc}^{+}$ state, and the ellipsis denotes contributions from higher resonances and the continuum. The thermal matrix element defining the meson decay constant is given by:
\begin{equation}\label{eq:matrixel}
\langle w | J_{\mu} | T_{cc}^{+}(q) \rangle = f_{T_{cc}^{+}}(T)\, m_{T_{cc}^{+}}(T)\, \varepsilon_{\mu},
\end{equation}
with $\varepsilon_{\mu}$ being the polarization vector. By substituting Eq.~(\ref{eq:matrixel}) into Eq.~(\ref{eq:phys1}), the scalar function 
$\Pi^{Phys}(q^2,T)$, which is the coefficient of the term 
containing $g_{\mu\nu}$ can be written:
\begin{equation}\label{eq:physgeneral}
\Pi^{Phys}(q^2,T) = \int_{0}^{\infty} \frac{\rho^{Phys}(s,T)}{s - q^{2}}\, ds,
\end{equation}
where the spectral density under the zero-width approximation reads:
\begin{align}
\rho^{Phys}(s)\big|_{{T_{cc}^{+}}} &= f_{T_{cc}^{+}}^{2}(T)\, m_{T_{cc}^{+}}^{2}(T)\,
\delta\!\left(s - m_{T_{cc}^{+}}^{2}(T)\right) \nonumber \\
&\quad + \theta\!\left(s - s_{0}(T)\right)\, \rho(s)\big|_{\text{PQCD}}. \label{eq:specden}
\end{align}
Here, the quantity $s_{0}(T)$ is the continuum threshold.

To construct the QCD side, we model the interpolating current $J_{\mu}(x)$ for the $T_{cc}^{+}$ as a compact axial-vector molecular configuration:
\begin{align}\label{curraxial}
J_\mu(x) &= \left[ \bar{u}_{a}(x)\gamma_5 c_{a}(x)\,\bar{d}_{b}(x)\gamma_\mu c_{b}(x) \right. \notag \\
         &\quad \left. - \bar{u}_{a}(x)\gamma_\mu c_{a}(x)\,\bar{d_{b}}(x)\gamma_5 c_{b}(x) \right],
\end{align}
where $a, b$ are colour indices. The interpolating current \( J_\mu(x) \) is constructed to carry the full physical signature of the studied meson, explicitly including its quark content and quantum numbers.

On the QCD side, the correlation function is evaluated by contracting 
the heavy- and light-quark fields using the corresponding thermal propagators. 
After performing the Wick contractions, the correlation function takes the form:
\begin{eqnarray}\label{eq:pi}
&&\Pi_{\mu\nu}^{\mathrm{QCD}}(q,T)=i\int d^{4}x\, e^{iq\cdot x}  \times \Big\{ \notag \\
&& \hspace{0.8em} \mathrm{Tr}[\gamma_{5}S_{c}^{aa^{\prime}}(x)\gamma_{5}S_{u}^{a^{\prime}a}(-x)]\mathrm{Tr}[\gamma_{\mu}S_{c}^{bb^{\prime}}(x)\gamma_{\nu}S_{d}^{b^{\prime}b}(-x)]\notag \\ 
&&-\mathrm{Tr}[\gamma_{5}S_{c}^{aa^{\prime}}(x)\gamma_{\nu}S_{u}^{a^{\prime}a}(-x)]\mathrm{Tr}[\gamma_{\mu}S_{c}^{bb^{\prime}}(x)\gamma_{5}S_{d}^{b^{\prime}b}(-x)]\notag \\ 
&&-\mathrm{Tr}[\gamma_{\mu}S_{c}^{aa^{\prime}}(x)\gamma_{5}S_{u}^{a^{\prime}a}(-x)]\mathrm{Tr}[\gamma_{5}S_{c}^{bb^{\prime}}(x)\gamma_{\nu}S_{d}^{b^{\prime}b}(-x)]\notag \\ 
&&+\mathrm{Tr}[\gamma_{\mu}S_{c}^{aa^{\prime}}(x)\gamma_{\nu}S_{u}^{a^{\prime}a}(-x)]\mathrm{Tr}[\gamma_{5}S_{c}^{bb^{\prime}}(x)\gamma_{5}S_{d}^{b^{\prime}b}(-x)] \Big\}\notag.\\ 
\end{eqnarray}
The thermal light quark propagator is given by
\cite{Mallik:1997kj}:
\begin{eqnarray}
S_{q}^{ij}(x) &=&i\frac{\slashed
x}{2\pi^{2}x^{4}}\delta_{ij}-\frac{
m_{q}}{4\pi^{2}x^{2}}\delta_{ij}  \notag \\
&-&\frac{\langle \bar{q}q\rangle }{12}\delta_{ij}-\frac{x^{2}}{192}%
m_{0}^{2}\langle \bar{q}q\rangle \Big[1-i\frac{m_{q}}{6}\slashed x \Big]%
\delta _{ij}  \notag \\
&+&\frac{i}{3}\Big[\slashed x \Big(\frac{m_{q}}{16}\langle
\bar{q}q\rangle
-\frac{1}{12}\langle u\Theta ^{f}u\rangle \Big)  \notag \\
&+&\frac{1}{3}\Big(u\cdot x\Big)\slashed u \langle u\Theta ^{f}u\rangle %
\Big]\delta _{ij}  \notag \\
&-&\frac{ig_{s}\lambda _{ij}^{A}}{32\pi ^{2}x^{2}}G_{A}^{\mu \nu }
\Big(\slashed x \sigma _{\mu \nu }+\sigma _{\mu \nu }\slashed
x\Big), \label{lightquarkpropagator}
\end{eqnarray}
where $u^\mu$ denotes the four-velocity of the heat bath and $\Theta^f_{\mu\nu}$ is the fermionic part of the energy-momentum tensor. The gluon condensate is defined as \cite{Mallik:1997pq}:
\begin{align}
\label{TrGG}
\langle \mathrm{Tr}^c\, 
G_{\alpha \beta } G_{\lambda \sigma} \rangle_T
&=
\left(g_{\alpha\lambda} g_{\beta\sigma}
- g_{\alpha\sigma} g_{\beta\lambda}\right) A
\nonumber \\
&
-\Big(
u_{\alpha} u_{\lambda} g_{\beta\sigma}
- u_{\alpha} u_{\sigma} g_{\beta\lambda}
\nonumber \\
&
- u_{\beta} u_{\lambda} g_{\alpha\sigma}
+ u_{\beta} u_{\sigma} g_{\alpha\lambda}
\Big) B,
\end{align}
where the coefficients $A$ and $B$ are given by
\begin{align}
\label{AB}
A &= \frac{1}{24}
\langle G^a_{\alpha \beta }
G^{a \alpha \beta}\rangle_T
+ \frac{1}{6}
\langle u^\alpha
\Theta^{g}_{\alpha \beta }
u^\beta\rangle_T ,
\nonumber \\
B &= \frac{1}{3}
\langle u^\alpha
\Theta^{g}_{\alpha \beta }
u^\beta\rangle_T .
\end{align}
The heavy quark propagator $S_Q^{ij}(x)$ $(\text{with}~Q = c)$ is given by:
\begin{eqnarray}\label{HeavyProp}
S_{Q}^{ij}(x)&=&i\int \frac{d^{4}k}{(2\pi )^{4}}e^{-ik\cdot x}\Bigg[ \frac{%
\delta _{ij}\Big( {\!\not\!{k}}+m_{Q}\Big)
}{k^{2}-m_{Q}^{2}}
\nonumber \\
&-& \frac{g_sG_{ij}^{\alpha \beta }}{4}\frac{\sigma _{\alpha \beta }\Big( {%
\!\not\!{k}}+m_{Q}\Big) +\Big(
{\!\not\!{k}}+m_{Q}\Big)\sigma_{\alpha
\beta }}{(k^{2}-m_{Q}^{2})^{2}}\nonumber \\
&+&\frac{g_s^{2}}{12}G_{\alpha \beta }^{A}G_{A}^{\alpha \beta
}\delta_{ij}m_{Q}\frac{k^{2}+m_{Q}{\!\not\!{k}}}{(k^{2}-m_{Q}^{2})^{4}}+\ldots\Bigg],
\end{eqnarray}
with $G_{A}^{\alpha \beta}$ denoting the external gluon field, $A=
1,2,...8$,~$\lambda_{ij}^{A}$ are the Gell-Mann matrices,
$t_{ij}^{A}=\lambda_{ij}^{A}/2$ with $G_{ij}^{\alpha\beta} \equiv G_A^{\alpha\beta} \lambda_{ij}^A/2$. 

Inserting the expressions from Eqs.~(\ref{lightquarkpropagator}) and (\ref{HeavyProp}) into Eq.~(\ref{eq:pi}) and decomposing the correlation function into Lorentz structures, we obtain: 
\begin{eqnarray}
\Pi_{\mu\nu}^{\mathrm{QCD}}(q,T) &=& \Pi_0^{\mathrm{QCD}}(q^2,T)\frac{q_\mu q_\nu}{q^2} \nonumber \\&+&\Pi_1^{\mathrm{QCD}}(q^2,T)(g_{\mu\nu}+\frac{q_\mu q_\nu}{q^2}) + \cdots,
\end{eqnarray} 
where $\Pi_0^{\mathrm{QCD}}(q^2,T)$ and $\Pi_1^{\mathrm{QCD}}(q^2,T)$ are invariant functions. Within the TQCDSR approach, the same Lorentz structure \( g_{\mu\nu} \) should be selected in the physical and QCD representations, \( \Pi_{\mu\nu}^{\text{Phys}}(q^{2}, T) \) and \( \Pi_{\mu\nu}^{\text{QCD}}(q^{2}, T) \), respectively. 
The QCD side of the correlation function is represented via a dispersion relation:
\begin{equation}
\Pi_1^{\mathrm{QCD}}(q^2,T) = \int_{\mathcal{M}^2}^{\infty} \frac{\rho^{\mathrm{QCD}}(s,T)}{s - q^{2}}\, ds,
\end{equation}
here $\mathcal{M}^2 = (m_u + m_d + 2m_c)^2$ and the spectral density is related to the imaginary part of the QCD correlator:
\begin{equation}\label{eq:rhoQCD}
\rho^{\mathrm{QCD}}(s,T) = \frac{1}{\pi} \, \mathrm{Im}[\Pi_1^{\mathrm{QCD}}(s,T)].
\end{equation}
 Note that, the above calculations is valid only in the zero-width approximation and thus cannot describe the temperature-dependent decay widths. To resolve this limitation, finite widths must be incorporated into all the expressions. This necessitates replacing the delta function in Eq.~\eqref{eq:specden} with the Breit-Wigner parameterization \cite{Dominguez:2012um,Dominguez:2007,Azizi:2010zza}:
\begin{widetext}
\begin{equation} \label{eq:Breit}
\delta\!\left(s - m_{{T_{cc}^{+}}}^{2}(T)\right) 
\;\;\longrightarrow\;\;
\text{const}  \times 
\frac{1}{\left[s - m_{{T_{cc}^{+}}}^{2}(T)\right]^{2} 
      + m_{{T_{cc}^{+}}}^{2}(T)\,\Gamma_{{T_{cc}^{+}}}^{2}(T)} \, .
\end{equation}
This observation reveals that hadronic species develop substantial thermal widths, $\Gamma_{{T_{cc}^{+}}}(T)$, at finite temperatures, which can be attributed to absorption in the thermal environment. The proportionality factor,
$ \text{const} = \frac{\,m_{T_{cc}^{+}}(T)\,\Gamma_{{T_{cc}^{+}}}(T)}{\pi}$, is established via integration throughout the range $(0,\infty)$. 

Having obtained the spectral representations on both the hadronic and QCD sides, we now equate the corresponding Lorentz structures under the assumption of quark–hadron duality. This procedure allows us to derive the basic sum rule expression for the $T_{cc}^{+}$ resonance.
Transferring the continuum contributions to the QCD side and applying the Borel transformation, the obtained formalism enables the numerical determination of the thermal hadronic parameters for the $T_{cc}^{+}$ resonance.

Taking into account the formalism of Eqs.~\eqref{eq:Breit} and (\ref{eq:specden}) in Eq.~(\ref{eq:physgeneral}), one subsequently obtains the sum rules describing the temperature evolution of mass, decay constant, and width:
\begin{equation}
\Pi_1^{\mathrm{QCD}}(q^2,T)=\Pi^{Phys}(q^2,T)=\frac{1}{\pi}\, f_{T_{cc}^{+}}^{2}(T)\, m_{{T_{cc}^{+}}}^{3}(T)\, \Gamma_{{T_{cc}^{+}}}(T) \int_{0}^{\infty} \frac{ds}{\left[s - m_{{T_{cc}^{+}}}^{2}(T)\right]^{2} + m_{{T_{cc}^{+}}}^{2}(T)\Gamma_{{T_{cc}^{+}}}^{2}(T)} \cdot \frac{1}{s - q^{2}}.
\end{equation}
This framework introduces three unknowns quantities: the temperature-dependent mass $m_{{T_{cc}^{+}}}(T)$, 
decay constant $f_{{T_{cc}^{+}}}(T)$, and width $\Gamma_{{T_{cc}^{+}}}(T)$. Applying the Borel transformation with respect to 
$q^{2}$ to both sides yields:
\begin{equation}\label{finitewidth1} \mathcal{B}_{q^{2}}\!\left[\Pi_1^{\mathrm{QCD}}(q^2,T)\right] =\frac{1}{\pi}\, f_{{T_{cc}^{+}}}^{2}(T)\, m_{{T_{cc}^{+}}}^{3}(T)\, \Gamma_{{T_{cc}^{+}}}(T) 
\int_{0}^{\infty} ds \,
\frac{e^{-s/M^{2}}}
     {\left[s - m_{{T_{cc}^{+}}}^{2}(T)\right]^{2} 
     + m_{{T_{cc}^{+}}}^{2}(T)\,\Gamma_{{T_{cc}^{+}}}^{2}(T)}.
\end{equation}
Two additional relations are required to determine these quantities. 
These can be obtained by applying the first and second derivatives 
with respect to $-1/M^{2}$ on both sides of Eq.~\eqref{finitewidth1}, namely:
\begin{equation}\label{finitewidth2}
\frac{1}{\pi}\, f_{{T_{cc}^{+}}}^{2}(T)\, m_{{T_{cc}^{+}}}^{3}(T)\, \Gamma_{{T_{cc}^{+}}}(T) 
\int_{0}^{\infty} ds \, 
\frac{s \, e^{-s/M^{2}}}
     {\left[s - m_{{T_{cc}^{+}}}^{2}(T)\right]^{2} 
     + m_{{T_{cc}^{+}}}^{2}(T)\,\Gamma_{{T_{cc}^{+}}}^{2}(T)} 
= \frac{-d}{d\!\left(1/M^{2}\right)} 
\left\{ \mathcal{B}_{q^{2}}\!\left[\Pi_1^{\mathrm{QCD}}(q^2,T)\right] \right\}.
\end{equation}
We conclude this section by examining the impact of finite-width effects on vacuum observables and on the system's thermal evolution. The inclusion of width effects modifies the core sum rules obtained from matching the physical and QCD representations in the Borel scheme, yielding the following expression:
\begin{equation}\label{finitewidth3}
\frac{1}{\pi}\, f_{{T_{cc}^{+}}}^{2}(T)\, m_{T_{cc}^{+}}^{3}(T)\, \Gamma_{T_{cc}^{+}}(T) 
\int_{0}^{\infty} ds \, 
\frac{s^{2} \, e^{-s/M^{2}}}
     {\left[s - m_{{T_{cc}^{+}}}^{2}(T)\right]^{2} 
     + m_{{T_{cc}^{+}}}^{2}(T)\,\Gamma_{{T_{cc}^{+}}}^{2}(T)} 
= \frac{d^{2}}{d \!\left(1/M^{2}\right)^{2}} 
\left\{ \mathcal{B}_{q^{2}}\!\left[\Pi_1^{\mathrm{QCD}}(q^2,T)\right] \right\}.
\end{equation}
The simultaneous solution of the coupled equations~\eqref{finitewidth1}--\eqref{finitewidth3} yields the thermal mass, decay constant, and width of the hadronic systems. The following section presents the numerical solutions to these equations and demonstrates their temperature dependence.
\end{widetext}

Also, we compute the hadronic parameters for the $T_{cc}^{+}$ resonance and, through heavy-quark substitution within the molecular picture, derive the properties of its bottom-sector counterpart, $T_{bb}^{+}$.
\section{Numerical Analysis}\label{sec:Numerical}
This section presents the thermal evolution of the mass, decay constant, and decay width, as determined from the QCDSR formalism. The QCDSR for the mass, decay constant, and decay width of the $T_{cc}^{+}$ state at finite temperature contains various quark, gluon, and mixed vacuum condensates as parameters. Their values are collected in Table~\ref{tab:Param}.

Beyond these parameters, we require the temperature-dependent quark
and gluon condensates, as well as the temperature-dependent energy
density. 

The temperature dependence of the light quark condensate is parametrized as
\begin{equation}
\langle \bar{q}q\rangle_{T}=\frac{\langle 0|\bar{q}q|0\rangle}{%
1+e^{18.10042(1.84692[\frac{1}{\mathrm{GeV}^{2}}]T_{\vphantom{2}}^{2}+4.99216[\frac{1}{\mathrm{GeV}}]T_{\vphantom{2}}-1)}}.
\end{equation}
This Fermi-Dirac-like suppression form, fitted to lattice QCD data \cite{Azizi:2019kzj, Ayala:2012ch,Bazavov:2019lgz, Cheng:2007jq}, captures the smooth crossover behavior across the deconfinement transition. At low temperatures, the condensate remains close to its vacuum value, while it vanishes exponentially at high temperatures, signaling chiral symmetry restoration. The rapid suppression occurs around $T_c \approx 155$ MeV, consistent with lattice QCD calculations.

For the gluonic and fermionic contributions to the energy density, we employ the parameterization derived in Ref.~\cite{Azizi:2019kzj}, which is based on the lattice QCD data presented in Ref.~\cite{Azizi:2016ddw,Cheng:2007jq}:
\begin{eqnarray}\label{tetamumu}
\langle \Theta _{00}^{g}\rangle &=&\langle \Theta^f
_{00}\rangle=\frac{1}{2}\langle \Theta _{00}\rangle \nonumber\\&=&T^{4}exp~{\Big(113.867\Big[\frac{1}{\mathrm{GeV}^{2}}\Big]T^{2}-12.190\Big[\frac{1}{\mathrm{GeV}}\Big]T
\Big)} \nonumber \\
&-& 10.141\Big[\frac{1}{\mathrm{GeV}}\Big]T^{5}.
\end{eqnarray}
We employ the thermal gluon condensate parametrization derived from QCD sum rules in conjunction with lattice QCD calculations, which provides a well-established description of non-perturbative effects in hot QCD matter~\cite{Ayala:2012ch}:
\begin{eqnarray}
\langle G^{2}\rangle  &=&\langle 0|G^{2}|0\rangle \Bigg[1-1.65\Big(\frac{T}{%
T_{c}}\Big)^{8.735}  \notag  \label{G2TLattice} \\
&+&0.04967\Big(\frac{T}{T_{c}}\Big)^{0.7211}\Bigg],
\end{eqnarray}%
with $\langle 0|G^{2}|0\rangle$ denoting the gluon condensate in vacuum.

The analytical structure of the sum rules manifests dependence on two auxiliary parameters: the vacuum continuum thresholds $s_{0}$ and the Borel mass parameters $M^{2}$. These threshold parameters are not freely adjustable but are physically constrained by the energy scale of the first excited states, which possess quantum numbers identical to those of the interpolating currents.
\begin{table}[h!]
\centering
\begin{tabular}{|c|c|}
\hline \hline
\textbf{Parameter} & \textbf{Value}~\cite{ParticleDataGroup:2024cfk,Shifman:1978bx,Reinders:1984sr,Narison:2010cg} \\ 
\hline \hline

$m_{c}$ 
& $1.28 \pm 0.03~\mathrm{GeV}$ \\

$m_{b}$ 
& $4.18^{+0.03}_{-0.02}~\mathrm{GeV}$ \\

$m_{u}$ 
& $2.16^{+0.49}_{-0.26}~\mathrm{MeV}$ \\

$m_{d}$ 
& $4.67^{+0.48}_{-0.17}~\mathrm{MeV}$ \\

$\langle \bar{q} q \rangle$ 
& $-(0.24 \pm 0.01)^3~\mathrm{GeV}^3$ \\

$\left\langle \alpha_s G^2 \right\rangle$ 
& $(0.020)~\mathrm{GeV}^4$ \\

$m_0^2$ 
& $(0.8 \pm 0.1)~\mathrm{GeV}^2$ \\

\hline \hline
\end{tabular}
\caption{Input QCD parameters used in the analysis.}
\label{tab:Param}
\end{table}\\

The thermal continuum threshold for the $T_{cc}^{+}(3875)$ state emerges as an additional auxiliary parameter requiring self-consistent determination. We incorporate the temperature-dependent continuum threshold parameterization following Ref.~\cite{Dominguez:2009mk,Dominguez:2010mx}:
\begin{eqnarray}  \label{s0T}
s_0(T)&=&s_0\left[1-\left(\frac{T}{T_c}\right)^8\right] +4m_c^2\left(\frac{T%
}{T_c}\right)^8,
\end{eqnarray}
here $s_0$ is the continuum threshold at $T=0$. 
This parameterization incorporates two essential physical features: (i) at low temperatures, $s_0(T)$ approaches the vacuum value $s_0$, reflecting the stability of hadronic resonances, and (ii) at the critical temperature, $s_{0}(T_{c}) \rightarrow 4m_{c}^{2}$, corresponding to the perturbative QCD threshold for open-charm production, which is consistent with the deconfinement. The power of eight in Eq.~(\ref{G2TLattice}) ensures a smooth transition between these two regimes and has been successfully employed in studies of heavy quarkonium and light mesons at finite temperatures.
Additionally,
\begin{itemize}
    \item \textbf{In light-quark systems} (e.g., $\rho$-meson):
    
    $s_0(T)/s_0(0) \to 0$ as $T \to T_c$, exhibiting a rapid approach to zero as $T$.
    
    \item \textbf{In heavy-light quark systems} (e.g., $D$-meson):
    
    $s_0(T)$ decreases gradually and approaches zero in the vicinity of $T_c$.
    
    \item \textbf{In heavy-heavy quark systems} (charmonium, bottomonium):
    
    $s_0(T)$ remains significantly above zero even beyond $T_c$; namely, these hadronic states maintain their existence above the critical temperature.
\end{itemize}

A temperature-dependent continuum threshold is applied here to doubly charmed exotic-molecular states. This parameter is physically constrained, and its value is dictated by the mass of the first excited state that shares the quantum numbers of the $T_{cc}^{+}$ interpolating current.
For $s_0$, we adopt the interval
\begin{equation}
17.5~\mathrm{GeV}^2\leq s_0\leq 19.5~\mathrm{GeV}^2,  \label{eq:19}
\end{equation}
over which the extracted physical observables exhibited only mild variations.

As the Borel parameters $M^{2}$ represent mathematical constructs rather than physical observables, the extracted quantities, such as masses, decay constants, and decay widths, must exhibit minimal dependence on their specific values. Consequently, we identify the optimal working regions for these parameters, such that the sensitivity of the sum rules to $M^{2}$ is substantially reduced. These regions are delineated by two complementary criteria: first, the contributions from continuum and excited states must remain subdominant; second, the influence of higher-dimensional operators must be sufficiently suppressed to ensure convergence of the operator product expansion.

Our results show that the perturbative term clearly dominates the OPE, and the ratio 
$\Pi(s_{0},M^{2}) / \Pi(\infty,M^{2})$ confirms good pole dominance and strong OPE convergence for 
$M^{2} \geq 2.8~\mathrm{GeV}^{2}$. 

\begin{figure*}[htbp]
    \centering    
    \includegraphics[width=0.48\linewidth]{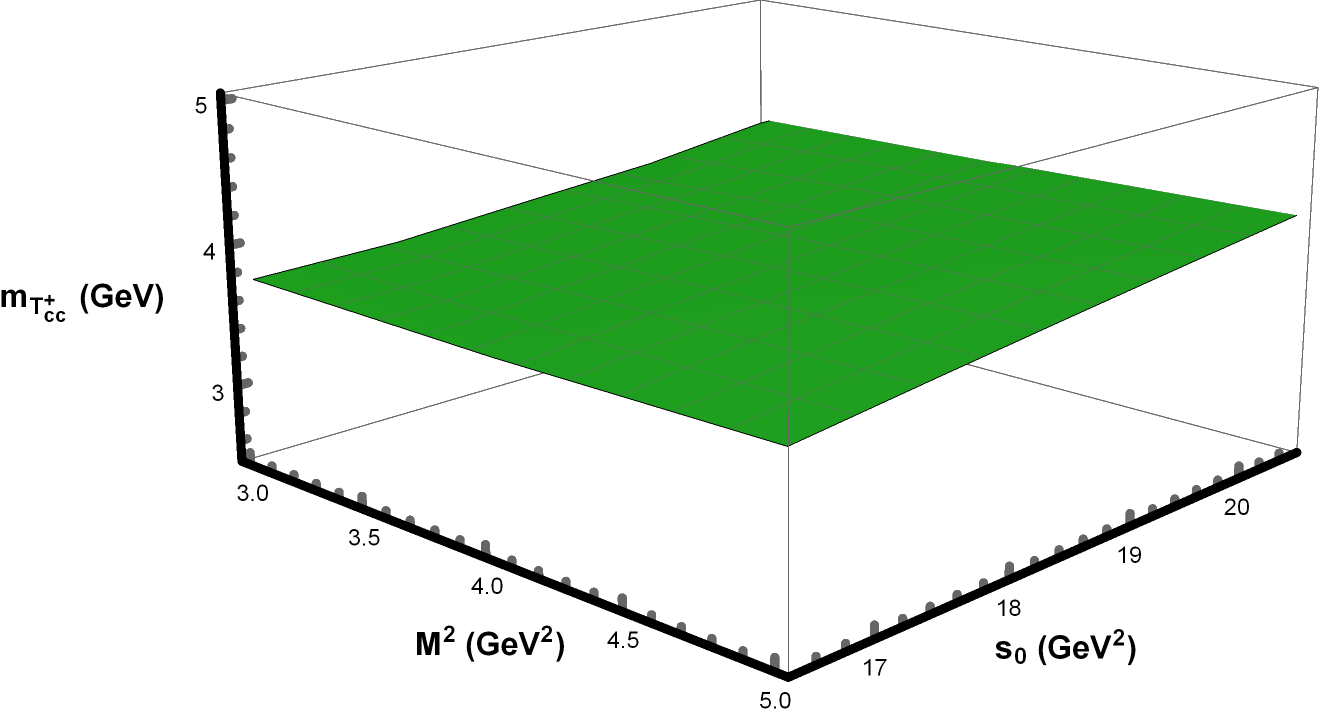} \hfill
    \includegraphics[width=0.48\linewidth]{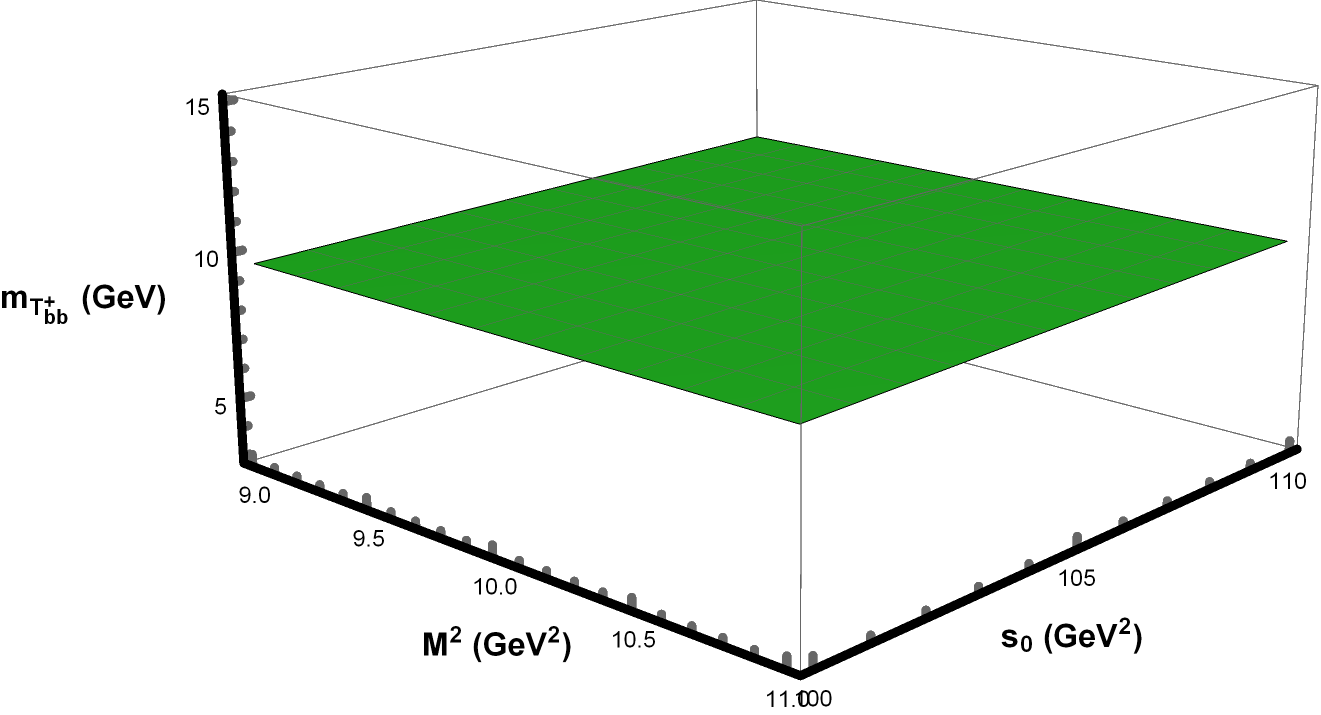}
    \caption{The vacuum mass of the hadronic molecule $T^{+}_{cc}$ (left panel) and its partner state $T_{bb}^{+}$ (right panel) as a function of the Borel parameter $M^{2}$ and the continuum threshold $s_{0}$.}
    \label{fig:MassTccandTbb}
\end{figure*}

\begin{figure*}[htbp]
    \centering
    \includegraphics[width=0.48\linewidth]{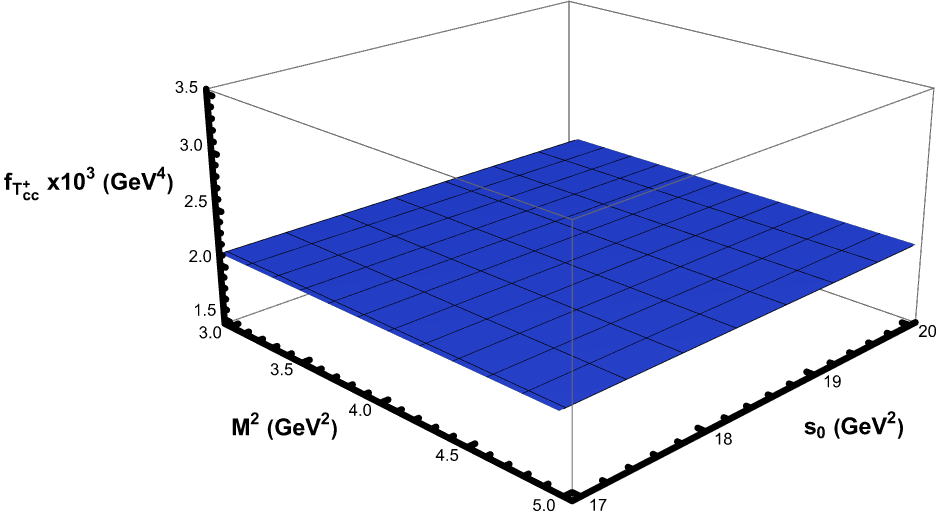} \hfill
    \includegraphics[width=0.48\linewidth]{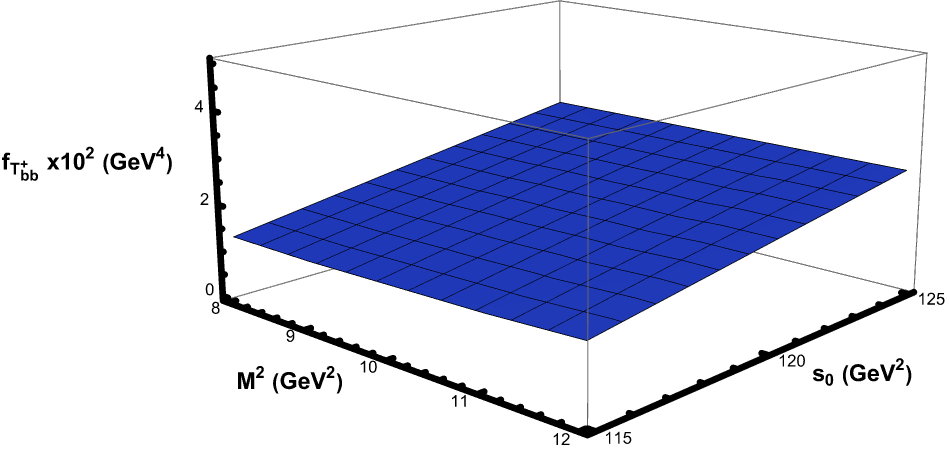}
    \caption{The vacuum decay constant of the hadronic molecule $T_{cc}^{+}$ (left panel) and its partner state $T_{bb}^{+}$ (right panel) as a function of the Borel parameter $M^{2}$ and the continuum threshold $s_{0}$.}
    \label{fig:ftccandTbb}
\end{figure*}

Accordingly, for the determination of mass, decay constant and width we notice the region of $ M^2$
\begin{equation}
3 ~\mathrm{GeV}^2\leq M^2\leq 5 ~\mathrm{GeV}^2,  \label{eq:M2}
\end{equation}
as safe for our goals.  The stability of the sum rules with respect 
to these model parameters is typically verified by plotting the relevant 
dependencies. As an illustration, we present in Figs.~\ref{fig:MassTccandTbb} and \ref{fig:ftccandTbb}
corresponding plot for the $T_{cc}^{+}$ and  $T_{bb}^{+}$ states.

In the end, the following results for the $T_{cc}^{+}$ and  $T_{bb}^{+}$ resonance in the $T=0$ limit of the TQCDSR model are shown in Table \ref{tab:TccResults} and \ref{tab:TbbResults}, which are consistent with the experimental and theoretical estimations reported in Refs.~\cite{LHCb:2021vvq,Navarra:2007yw,Ferah:2025jle}.
\vspace{1mm}\\
\begin{table}[h!]
\centering
\caption{Mass, decay constant, and width of the $T_{cc}^+$ state at $T=0$ 
from different theoretical approaches and experimental measurements.}
\begin{tabular}{lccc}
\hline\hline
 & $m_{T_{cc}^+}$ (~\text{MeV}) & $f_{T_{cc}^+}$$\times 10^{3}$ ~(\text{GeV}$^4$) & $\Gamma_{T_{cc}^+}$ (~\text{keV}) \\
\hline
\multicolumn{4}{l}{\textit{Molecular Structure}} \\
\hline
This work & $3961.17 \pm 78.5$ & $2.1 \pm 0.4$ & $434.95 \pm 7.66$ \\
Exp.~\cite{LHCb:2021vvq} & $3874.74 \pm 0.1$ &  $- $ & $410 \pm 165$ \\
Ref.~\cite{Agaev:2022ast} & $4060 \pm 130$ & $5.1 \pm 0.8$ & $3.8 \pm 1.7 \text{ MeV}$ \\
Ref.~\cite{Chen:2021vhg} & $3876$ & $- $& $412$ \\

Ref.~\cite{Ren:2021dsi} & $3875.1$ & $-$ & $-$ \\
\hline
\multicolumn{4}{l}{\textit{Tetraquark Structure}} \\
\hline
Ref.~\cite{Ferah:2025jle} & $3887.3 \pm 143$ &  $- $ &  $- $ \\
Ref.~\cite{Agaev:2021vur} & $3868 \pm 124$ & $5.03 \pm 0.79$ & $489 \pm 92$ \\
Ref.~\cite{Wang:2017dtg} & $3900 \pm 90$ &  $- $ &  $- $ \\
Ref.~\cite{Braaten:2020nwp} & $3947 \pm 11$ &  $- $ &  $- $ \\
\hline\hline
\end{tabular}
\label{tab:TccResults}
\end{table}
\begin{table}[h!]
\centering
\caption{Mass, decay constant, and width of the $T_{bb}^+$ state at $T=0$ 
from different theoretical approaches.}
\begin{tabular}{lccc}
\hline\hline
 & $m_{T_{bb}^+}$ (~\text{MeV}) & $f_{T_{bb}^+}$ $\times10^{2}$ ~(\text{GeV}$^4$) & $\Gamma_{T_{bb}^+}\times 10^{5} (~\text{keV})$ \\
\hline
\multicolumn{4}{l}{\textit{Molecular Structure}} \\
\hline
This work & $10358 \pm 280$ & $1.95 \pm 0.71 $ & $1.74 \pm 0.09$  \\
Ref.~\cite{Ren:2021dsi} & $10598$ & $-$ & $-$ \\
\hline
\multicolumn{4}{l}{\textit{Tetraquark Structure}} \\
\hline
Ref.~\cite{Leskovec:2019ioa} & $10476 \pm 24$ & $-$ & $-$ \\
Ref.~\cite{Karliner:2017qjm} & $10389.4 \pm 12$ & $-$ & $-$ \\
Ref.~\cite{Bicudo:2016ooe} & $10545 \pm 38$ & $-$ & $-$ \\
Ref.~\cite{Agaev:2018khe} & $10035 \pm 260$ & $1.38 \pm 0.27$ & $7.17\pm1.23$ \\
\hline\hline
\end{tabular}
\label{tab:TbbResults}
\end{table}\\

\begin{figure*}[htbp]
    \centering
    \includegraphics[width=0.45\linewidth]{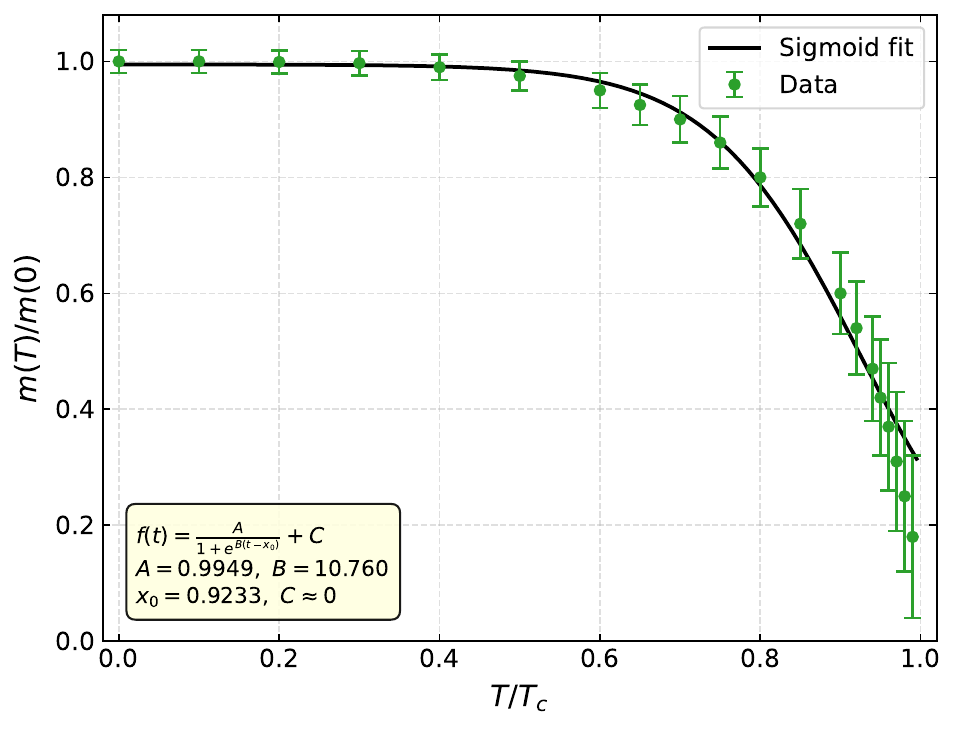} \hfill
    \includegraphics[width=0.45\linewidth]{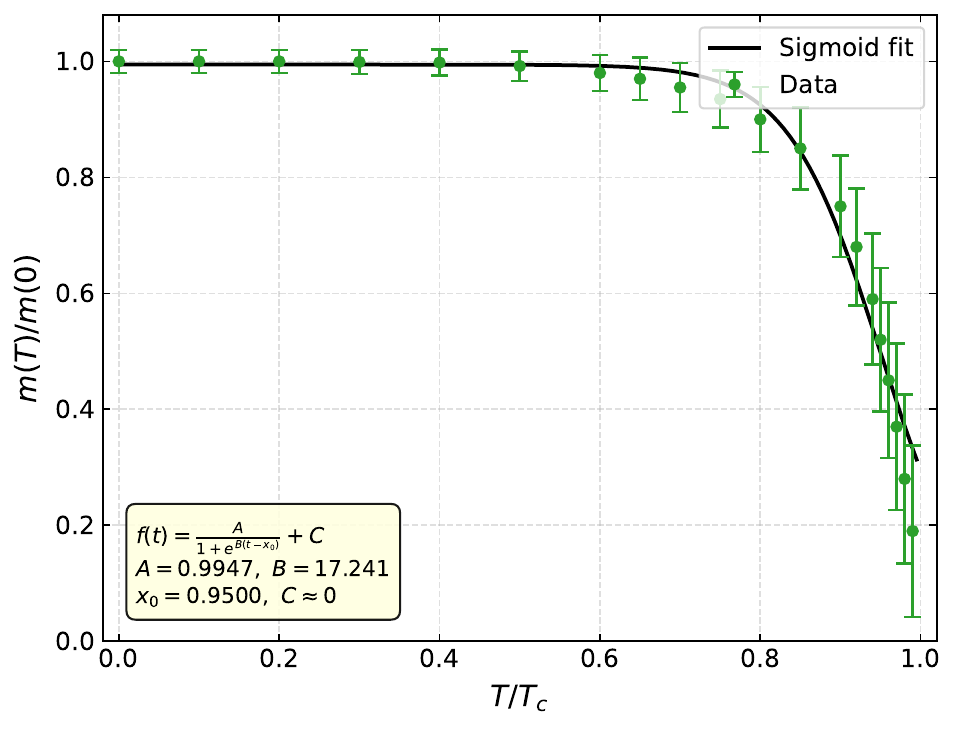}
    \caption{Thermal evolution of the mass of the $T_{cc}^{+}$ state (left panel) and its heavy-flavor partner $T_{bb}^{+}$ (right panel) within the molecular picture, evaluated for fixed continuum threshold parameters $s_{0}$.}
    \label{fig:MassTccandTbbattemperature}
\end{figure*}

\begin{figure*}[htbp]
    \centering
    \includegraphics[width=0.45\linewidth]{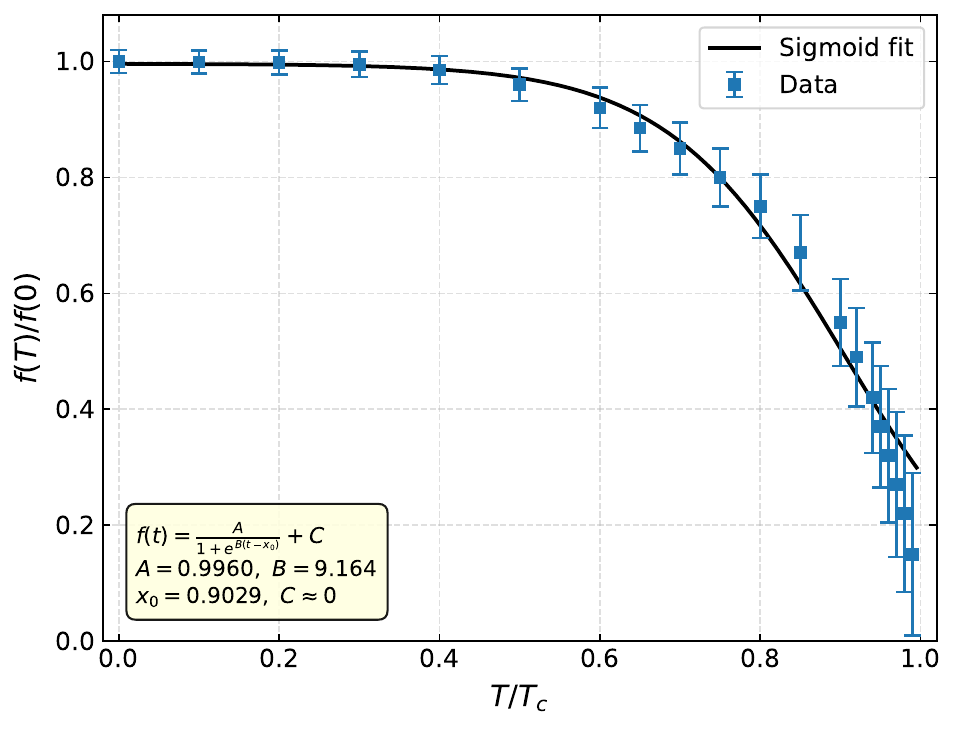} \hfill
    \includegraphics[width=0.45\linewidth]{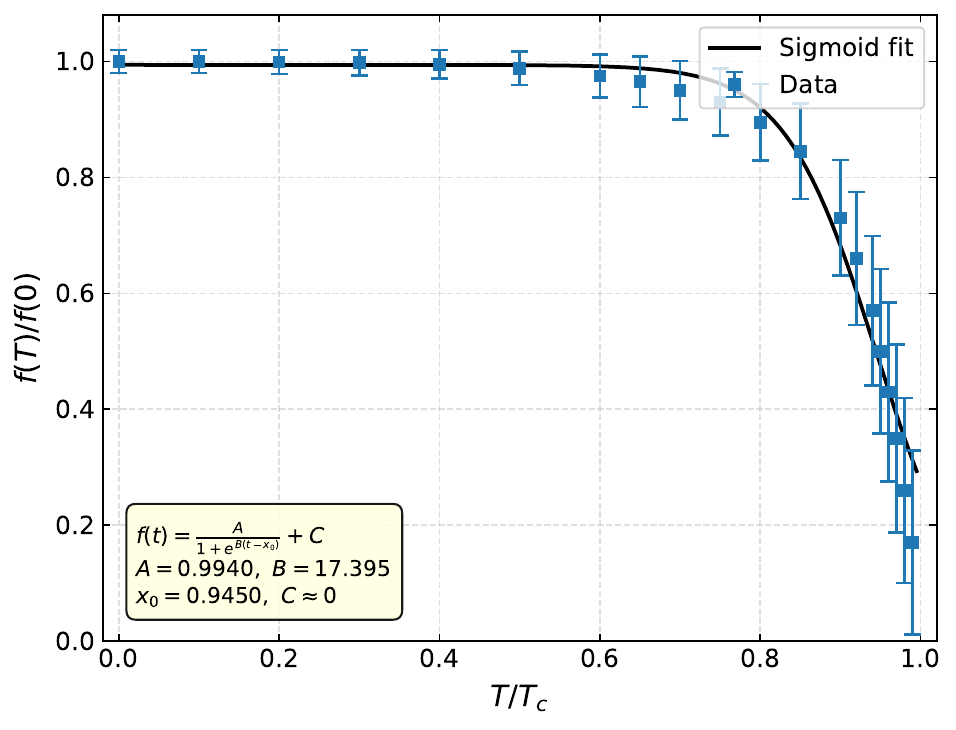}
    \caption{Thermal behavior of the decay constant for the $T_{cc}^{+}$ state (left panel) and its $T_{bb}^{+}$ partner (right panel) in the molecular model framework, computed using fixed continuum threshold parameters $s_{0}$.}
    \label{fig:ftccandTbbattemperature}
\end{figure*}

\begin{figure*}[htbp]
    \centering
    \includegraphics[width=0.45\linewidth]{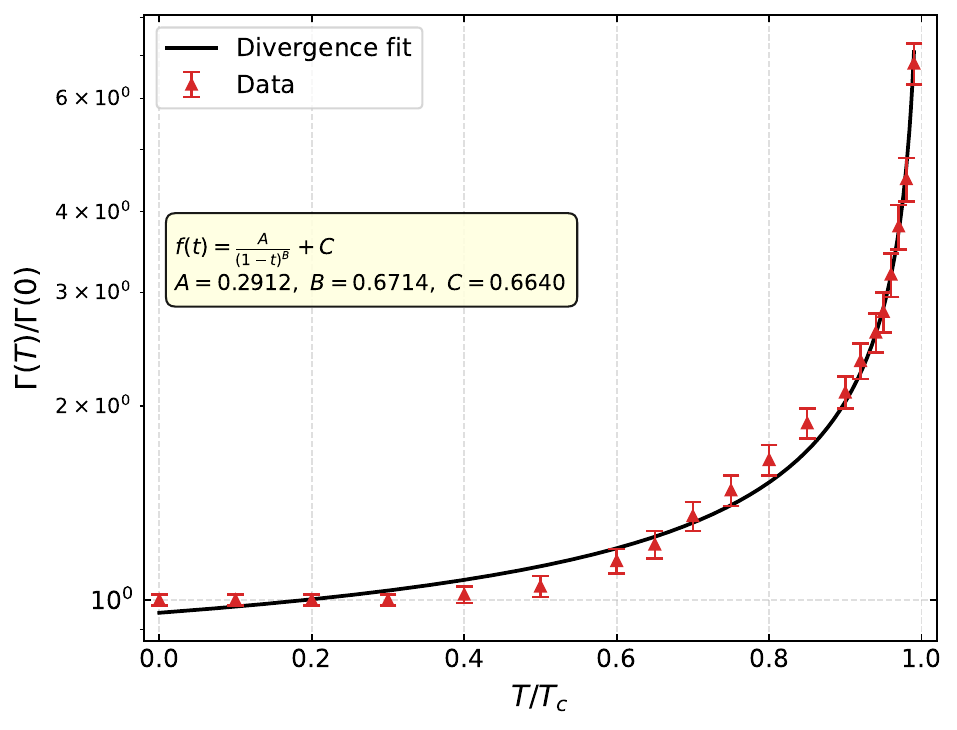} \hfill
    \includegraphics[width=0.45\linewidth]{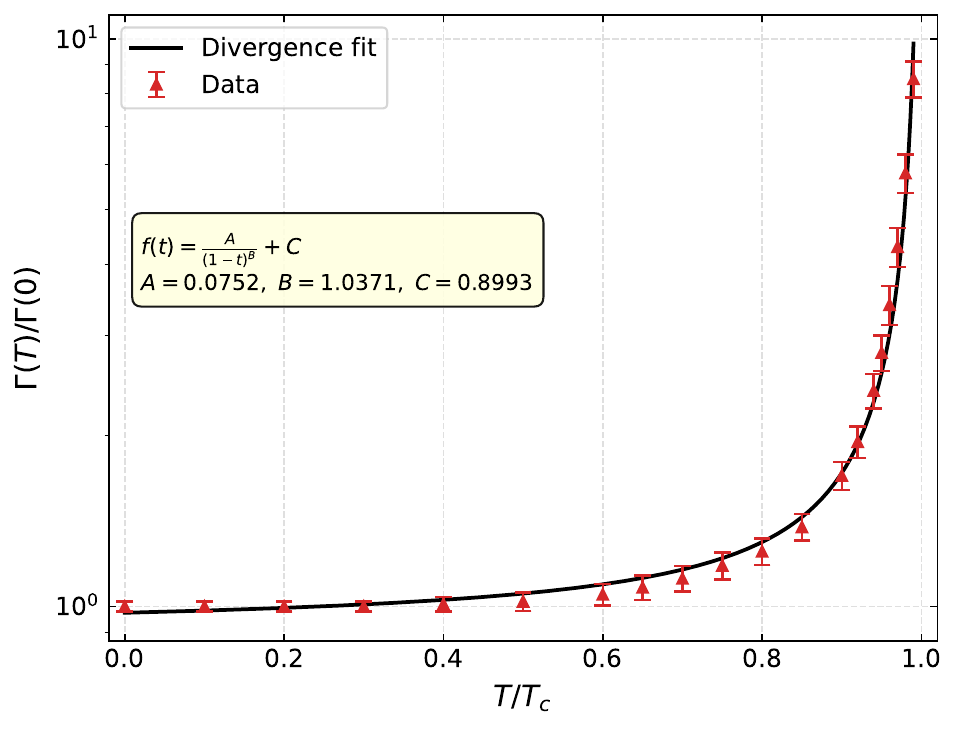}
    \caption{Thermal evolution of the decay width for the $T_{cc}^{+}$ state (left panel) and its $T_{bb}^{+}$ partner (right panel) within the molecular picture.}
    \label{fig:gammaTccandTbb}
\end{figure*}

Next, we investigate the temperature dependence of the masses, 
decay constants and widths of the $T_{cc}^{+}$ and  $T_{bb}^{+}$ states. For this purpose, their variations are plotted as functions of temperature under the molecular assumption, as shown in Figs.~\ref{fig:MassTccandTbbattemperature}, \ref{fig:ftccandTbbattemperature} and \ref{fig:gammaTccandTbb}, respectively. In contrast to the masses and decay constants, the widths increase with temperature.

To complete the analysis, we investigate how the width varies
with temperature. The dependence of $ \Gamma_{ T_{cc}^{+}}( T)$ and $\Gamma_{ T_{bb}^{+}}( T)$
on $ T$ is plotted in Fig.~\ref{fig:gammaTccandTbb}, revealing a pronounced increase that becomes steep near the critical region.
Our result for $\Gamma_{T_{cc}^{+}}$ is consistent with the LHCb Collaboration's measured value \cite{LHCb:2021vvq}. 
As the system approaches the critical temperature for deconfinement \(( T \approx T_{c})\), the hadronic resonances gradually disappear, since their decay widths become exceedingly large. At the same time, their couplings to the interpolating currents vanish.

\section{Conclusions}\label{sec:conclusions}
In the present investigation, we conducted a systematic analysis of the $T_{cc}^{+}$ exotic meson within the hadronic molecular picture, employing the TQCDSR framework to compute its temperature-dependent spectroscopic parameters. The derived thermal sum rules provide a consistent framework for quantifying the medium-induced modifications to the mass, decay constant, and width of the $T_{cc}^{+}$ resonance.

Our results establish that the mass, decay constant, and width exhibit remarkable thermal stability, maintaining their vacuum values up to 
$T \simeq 120\,\mathrm{MeV}$; beyond this point, the mass and decay constant decrease with increasing temperature, whereas the width increases dramatically.
At the deconfinement temperature, the decay constant decreases to about 
$25\%$ of its vacuum value, and the mass is reduced to roughly 
$28\%$. 
The thermal width of the $T_{cc}^{+}$ state exhibits a sharp enhancement as 
$T$ approaches $T_c$, reflecting strong thermal suppression of the pole contribution and the onset of hadronic dissolution in the medium. 
Near $T_c$, the width increases by approximately a factor of 6.

For the $T_{bb}^{+}$ state, the mass decreases to roughly 
$20\%$ of its vacuum value, while the decay constant diminishes to about 
$18\%$. 
This behavior clearly demonstrates that elevated temperatures significantly alter the internal structure and binding strength of these states, underscoring the sensitivity of their multiquark nature to the thermal medium. 
Correspondingly, the thermal width of the $T_{bb}^{+}$ state also shows a marked rise near 
$T_c$, increasing by approximately a factor of 10.

 For temperatures well below the critical point, $T \ll T_c$, both the mass, decay constant, and decay width remain nearly unchanged, reflecting the stability of the hadronic state against thermal effects. However, as $T \lesssim T_c$, the suppression of the quark and gluon condensates together with the reduction of the continuum threshold leads to a rapid loss of pole dominance. Consequently, the extracted decay width exhibits a sharp increase near the critical temperature, which should be interpreted as a signal of the dissolution of the $T_{bb}^+$ state rather than a genuine physical broadening.

In a hot and dense medium, the fundamental properties of hadrons such as mass, decay width, and decay constant undergo significant modifications due to medium-induced effects. 
The reduction of the pole mass with increasing temperature reflects the partial restoration of chiral symmetry and the weakening of quark-gluon binding. 
The decay width generally broadens at high temperatures, indicating an enhanced interaction rate between the hadron and the surrounding thermal particles, which leads to a shorter hadron lifetime. 
The decay constant, on the other hand, tends to decrease as the temperature approaches the deconfinement region, signifying the gradual dissociation of the hadronic bound state into its quark constituents. These temperature-dependent variations provide essential insights into the in-medium behavior of hadrons and the transition from the hadronic phase to the quark-gluon plasma.

The simultaneous suppression of the mass and decay constant,
accompanied by a sharp enhancement of the decay width near
$T_c$, constitutes a coherent and distinctive signature of
deconfinement and the onset of the quark-gluon plasma phase
transition. Furthermore, the derived temperature dependence provides a quantitative framework for interpreting the observables in heavy-ion collision experiments. Our theoretical predictions for thermal evolution can be directly tested against upcoming measurements from LHC heavy-ion programs, thereby making finite-temperature QCD effects experimentally accessible.

Future in-medium investigations will facilitate rigorous examination of hadronic properties under extreme temperature and density conditions. Confronting forthcoming experimental data with phenomenological calculations will elucidate the nature of exotic mesonic states and yield crucial insights into their internal architecture and quark–gluon dynamics. Such comparative analyses will advance our understanding of non-perturbative QCD phenomena in hot and dense environments.
\section*{Acknowledgments}
The authors are grateful to K. Azizi and H. Sundu for their valuable contributions and helpful suggestions.
%

\end{document}